\documentclass[amsmath,amssymb,floatfix,aps,prl,reprint,
               twocolumn,superscriptaddress
               ]{revtex4-2}

\usepackage{graphicx}% Include figure files
\usepackage{dcolumn}% Align table columns on decimal point
\usepackage{bm}% bold math
\usepackage{amsmath}
\usepackage[colorlinks=true, allcolors=blue]{hyperref}

\graphicspath{ {./} }

\begin{document}

\title{Facile \textit{ab initio} approach for self-localized polarons from canonical transformations}

\author{Nien-En Lee}
\affiliation{Department of Applied Physics and Materials Science, California Institute of Technology, Pasadena, California 91125, USA}
\affiliation{Department of Physics, California Institute of Technology, Pasadena, California 91125, USA}

\author{Hsiao-Yi Chen}
\affiliation{Department of Applied Physics and Materials Science, California Institute of Technology, Pasadena, California 91125, USA}
\affiliation{Department of Physics, California Institute of Technology, Pasadena, California 91125, USA}

\author{Jin-Jian Zhou}
\affiliation{Department of Applied Physics and Materials Science, California Institute of Technology, Pasadena, California 91125, USA}

\author{Marco Bernardi}
\affiliation{Department of Applied Physics and Materials Science, California Institute of Technology, Pasadena, California 91125, USA}

%\date{\today}

\begin{abstract}
Electronic states in a crystal can localize due to strong electron-phonon ($e$-ph) interactions, forming so-called small polarons. 
Methods to predict the formation and energetics of small polarons are either computationally costly or not geared toward quantitative predictions. 
Here we show a formalism based on canonical transformations to compute the polaron formation energy and wavefunction using \textit{ab initio} $e$-ph interactions. 
Comparison of the calculated polaron and band edge energies allows us to determine whether charge carriers in a material favor a localized small polaron over a delocalized Bloch state. 
Due to its low computational cost, our approach enables efficient studies of the formation and energetics of small polarons, as we demonstrate by investigating electron and hole polaron formation in alkali halides and metal oxides and peroxides. We outline refinements of our scheme and extensions to compute transport in the polaron hopping regime.
\end{abstract}

%\keywords{}

\maketitle
Self-localized (small) polarons are charge carriers that interact strongly with the lattice vibrations, becoming trapped as a result of the local lattice distortion~\cite{Emin1982}. Small polarons are essential to understanding electrical transport and optical properties in a wide range of materials, including transition metal oxides, alkali halides and organic molecular crystals \cite{Lany2015, Castner1957, Schein1978, Tim-PRX}. The presence of small polarons in these materials is typically associated with a diffusive, thermally activated charge transport regime characterized by low mobility values, typically less than $1\; \textrm{cm}^2/\textrm{V}\,\textrm{s}$ \cite{Tuller1977}. Recent progress has enabled direct observation of small polaron states \cite{Sezen2015, Pastor2019} and clarified their important role in various technologies \cite{Kang2012, Ong2012, Wu2015, Cortecchia2017, Salamon2001}.\\
\indent
The theoretical treatment of small polarons was pioneered by Holstein \cite{Holstein1959} based on ideas from Landau and Pekar \cite{Rashba2015}. It was later extended by Lang and Firsov \cite{Lang1962}, and generalized by Munn and Silbey \cite{Munn1980, Munn1985} and Hannewald et al. \cite{Hannewald200401} to improve the description of electron-phonon ($e$-ph) interactions. The resulting small polaron theory can qualitatively demonstrate the transition from bandlike to hopping transport observed in experiments \cite{Schein1978, Bogomolov1967, Bottger1976}. Yet, the presence of a self-localized polaron state is typically assumed in these theories rather than directly predicted, and most theoretical treatments of polarons are not geared toward quantitative predictions on real materials as they rely on empirical parameters and take into account only one or a few vibrational modes.\\ 
\indent
Early work formulated the problem of polaron formation as a competition of energies for localizing an electronic state, which relaxes the lattice, but increases the electron kinetic energy~\cite{Emin1972}. 
Despite this intuition, whether charge carriers form small polarons or not remains controversial in many materials. For example, photoemission experiments found no evidence of small polarons in SrTiO$_3$ \cite{Meevasana2010} although mobility and optical measurements suggested their presence \cite{Keroack1984, Crespillo2018}.\\
\indent
First-principles calculations can accurately compute the electronic structure, lattice dynamics and $e$-ph coupling~\cite{Martin2004}, and are ideally suited to provide quantitative approaches for treating both large and small polarons. However, existing studies have focused on semiconductors and insulators without small polaron effects \cite{Kas2014, Zhou2016, Lee2018, Zhou2018, Zhou2019, Kang2019, Lee2020}. First-principles calculations of small polarons involve supercells with excess charge or defects explicitly added \cite{Janotti2014, Kokott2018, Yuan2019, Tsunoda2019}. While useful, these approaches require computationally costly calculations with many atoms, and their reliability is limited by the accuracy of density functional theory (DFT) exchange-correlation functionals and the treatment of charged systems in DFT. 
%Recently proposed methods \citep{Sio201901, Sio201902} formulate the polaron problem in reciprocal space, leading to computationally expensive calculations which, as we discuss below, may not guarantee polaron self-localization.  
A rigorous and convenient first-principles approach connecting standard small polaron theory~\cite{Devreese2000} and modern \textit{ab initio} $e$-ph calculations would be expedient.\\
%
% HERE WE SHOW
%
\indent
Here we show an efficient approach to compute the small polaron energy in a localized basis starting from a trial polaron wavefunction. 
Employing a canonical transformation formalism \cite{Holstein1959}, we construct a self-localized polaron state that is free from hopping and decoupled from all vibrational modes~\footnote{Note that here we define a small polaron as a self-trapped electronic state, regardless of its spatial extent.}. We determine whether an electron or hole charge carrier self-localizes by comparing the energy of the polaron state with the conduction or valence band edge, thus predicting whether a small polaron forms and determining its formation energy. 
The computational cost of our scheme is equivalent to a DFT calculation on a unit cell plus an inexpensive $e$-ph computational step. 
Its efficiency allows us to investigate small polarons in various alkali halides, oxides and perovskites with minimal computational effort. 
Our work bridges the gap between standard small polaron theory and modern \textit{ab initio} $e$-ph calculations.\\%, formulating an efficient computational approach to treat small polarons from first principles.\\
% t, and a large polaron as a delocalized polaron state whose mobility exhibits a bandlike power-law temperature dependence
%
%
\indent
We derive the effective small polaron Hamiltonian in a distorted lattice through a canonical transformation~\cite{Holstein1959}, inspired by the treatment of the charged harmonic oscillator (CHO) in an external electric field $E$~\cite{Mahan2000}. %(see the companion paper for detailed derivations). 
The Hamiltonian of a one-dimensional CHO is
%(see the Supplemental Material~\cite{SupplementalMaterial} for additional derivations). The Hamiltonian of a one-dimensional CHO is
%
\begin{align}
\label{H_CHO}
H^{\textrm{(CHO)}} = \omega ( b^{\dagger} b + \frac{1}{2} )
      + \omega g ( b + b^{\dagger} ),
\end{align}
where $b^{\dagger}$ and $b$ are oscillator creation and annihilation operators, and the coupling parameter is $g=eE/\sqrt{2m\omega^3}$, with $e$, $m$, $\omega$ the charge, mass and frequency of the oscillator, respectively. Here and below we set $\hbar=1$. To solve the CHO Hamiltonian, the common approach is to stretch the oscillator spring to its new equilibrium position using the canonical transformation of operators $\mathcal{O} \rightarrow \widetilde{\mathcal{O}} = e^S \mathcal{O}e^{-S}$. Defining the CHO generator as $S^{\textrm{(CHO)}} =  g(b^{\dagger}-b)$, this transformation gives
\begin{gather}
\widetilde{b}=b-g,
\label{transformed_b_CHO}\\
\widetilde{H}^{\textrm{(CHO)}}=\omega(b^{\dagger}b+\frac{1}{2}) - \omega g^2.
\label{transformed_H_CHO}
\end{gather}
The shift of the operator $b$ in Eq.~(\ref{transformed_b_CHO}) amounts to shifting the coordinate system:
\begin{align}
\label{CHO_x_tilde}
\widetilde{x}=\frac{1}{\sqrt{2m\omega}}(\widetilde{b}^{\dagger}+\widetilde{b})=\frac{1}{\sqrt{2m\omega}}(b^{\dagger}+b) + x_0,
\end{align}
where $x$ is the position operator and $x_0=-2g/\sqrt{2m\omega}$ is the new equilibrium position. The second term in Eq.~(\ref{transformed_H_CHO}) is always negative and can be interpreted as the energy decrease resulting from relaxing the oscillator to a new equilibrium position due to the electrical force, because $- \omega g^2=\frac{1}{2}m \omega^2 {x_0}^2 + eEx_0$.\\
\indent
Inspired by Holstein's treatment~\cite{Holstein1959}, we perform an analogous transformation on the $e$-ph Hamiltonian in the electronic Wannier \cite{Marzari2012} and phonon momentum basis,
\begin{align}
H = & \sum_{mn} \varepsilon_{mn} a^{\dagger}_{m}a_{n} 
      + \sum_{\textbf{Q}} \omega_{\textbf{Q}}
      \left(b^{\dagger}_{\textbf{Q}}b_{\textbf{Q}} 
      + \frac{1}{2}\right)\\
    & +  \frac{1}{ \sqrt{N_{\Omega}} }
      \sum_{mn}\sum_{\textbf{Q}}\omega_{\textbf{Q}}
      g_{\textbf{Q}mn}
      \left(b^{\dagger}_{\textbf{Q}}+b_{-\textbf{Q}}\right)
      a^{\dagger}_{m}a_{n}\nonumber.
\end{align}
Here, $n=j_n \textbf{R}_{n}$ is a collective index labelling the $j_n$-th Wannier function (WF) in the unit cell with origin at the Bravais lattice vector $\textbf{R}_{n}$, while $a_n = a_{j_n \textbf{R}_{n}}$ is the corresponding electron annihilation operator and $b_{\textbf{Q}}$ is the phonon annihilation operator, where $\textbf{Q}$ is a collective label for the phonon mode $\nu$ and momentum $\textbf{q}$. The hopping strength and phonon energy are denoted as $\varepsilon_{mn}$ and $\omega_{\textbf{Q}}$, respectively, and $N_{\Omega}$ is the number of unit cells in the crystal. 
The $e$-ph coupling matrix element in the Wannier basis, denoted as $g_{\textbf{Q}mn}$, does not include the phonon frequency factor, different from the standard convention \cite{Perturbo2020}. Also recall that the $e$-ph coupling needs to satisfy the relation $g^{*}_{\textbf{Q}mn}= g_{-\textbf{Q}nm}$ for the Hamiltonian to be Hermitian.\\
\indent
We define the generator $S$ as
\begin{gather}
S = \sum_{mn}C_{mn}a^{\dagger}_{m}a_{n},\\
\label{C_mn_polaron}
C_{mn} =  \frac{1}{ \sqrt{N_{\Omega}} } \sum_{\textbf{Q}} B_{\textbf{Q}mn}
    (b^{\dagger}_{\textbf{Q}}-b_{-\textbf{Q}}),
\end{gather}
and using the transformation $\mathcal{O} \rightarrow \widetilde{\mathcal{O}} = e^S \mathcal{O}e^{-S}$ we obtain the transformed electron and phonon annihilation operators, respectively, as
\begin{align}
\label{a_tilde}
\widetilde{a}_m &= \sum_{n} e^{-C}_{mn}a_n,\\
\label{b_tilde}
\widetilde{b}_{\textbf{Q}} &= b_{\textbf{Q}}
    - \frac{1}{ \sqrt{N_{\Omega}} } \sum_{mn}B_{\textbf{Q}mn}a^{\dagger}_{m}a_n,
\end{align}
where $e^{-C}_{mn}$ is shorthand for the phonon operator
\begin{align}
e^{-C}_{mn} = 
    \delta_{mn} - C_{mn} 
    + \frac{1}{2!}\sum_{i}C_{mi}C_{in}-\cdots,
\end{align}
with $C_{mn}$ defined in Eq.~(\ref{C_mn_polaron}). Above, we introduced the undetermined distortion coefficients $B_{\textbf{Q}mn}$ which, analogous to the coupling $g$ in the CHO example, quantify how the transformation stretches the spring of each phonon mode to a new equilibrium position due to the electrical forces applied on the lattice by the charge carrier. 
This physical interpretation is manifest in Eq.~(\ref{b_tilde}), where one changes the basis to a distorted lattice configuration in analogy with Eq.~(\ref{transformed_b_CHO}), implying that the operators $a^{\dagger}_n$ and $b^{\dagger}_\textbf{Q}$ create a polaron or phonon, respectively, in the distorted lattice. To make the transformation unitary, the distortion coefficients need to satisfy $B^{*}_{\textbf{Q}mn}= B_{-\textbf{Q}nm}$, so that the operators $C_{mn}$ and $S$ are both anti-Hermitian.\\
\indent
The polaron Hamiltonian is obtained by substituting the transformed electron and phonon operators~\footnote{In deriving the polaron Hamiltonian, we assumed that the carrier concentration is low enough that polaron-polaron interactions can be neglected.}:
\begin{align}
\label{H_tilde_ta}
\widetilde{H}=& \sum_{mn} E_{mn} a^{\dagger}_{m} a_n
          + \sum_{\textbf{Q}} \omega_{\textbf{Q}}
             (b^{\dagger}_{\textbf{Q}}b_{\textbf{Q}}+\frac{1}{2})
             \\
          &+ \frac{1}{ \sqrt{N_{\Omega}} }
             \sum_{mn\textbf{Q}}\omega_{\textbf{Q}}
             G_{\textbf{Q}mn}
             (b^{\dagger}_{\textbf{Q}} + b_{-\textbf{Q}})
             a^{\dagger}_{m}a_n,\nonumber
\end{align}
where the polaron hopping strength $E_{mn}$ and the residual polaron-phonon (pl-ph) coupling constant $G_{\textbf{Q}mn}$ are defined respectively as
\begin{gather} 
E_{mn} = \langle \widetilde{\varepsilon}\rangle_{mn}  +
     \frac{1}{N_{\Omega}}\sum_{i\textbf{Q}} \omega_{\textbf{Q}} B_{-\textbf{Q}mi}
     \left( B_{\textbf{Q}in} - 2 \langle \widetilde{g}_{\textbf{Q}} \rangle_{in} \right),
     \nonumber\\
\label{polaron_parameters}
G_{\textbf{Q}mn} = \langle \widetilde{g}_{\textbf{Q}} \rangle_{mn} - B_{\textbf{Q}mn},
\end{gather}
and the angle brackets $\langle\cdots\rangle$ indicate a thermal average over phonon states. In this effective polaron Hamiltonian, the transformed hopping and $e$-ph coupling matrices $\widetilde{\varepsilon}_{mn}$ and $\widetilde{g}_{\textbf{Q}mn}$, denoted as $\widetilde{M}_{mn}$, are defined as
\begin{align}
\label{M_tilde}
\widetilde{M}_{mn} = \sum_{ij}e^{C}_{mi}M_{ij}e^{-C}_{jn}.
\end{align}
These transformed matrices still contain phonon operators (through the operator $C_{mn}$). Following Holstein \cite{Holstein1959}, we take their thermal average in Eq.~(\ref{polaron_parameters}) 
to obtain the effective polaron Hamiltonian in Eq.~(\ref{H_tilde_ta}).\\
\indent
We then set the distortion coefficients to
\begin{align}
\label{ansatz}
B_{\textbf{Q}mn} =  g_{\textbf{Q}mn} \delta_{mn},
\end{align}
and show that this choice leads to a self-localized polaron state. Using this ansatz, the thermal average of the transformed matrix can be written as $\langle \widetilde{M}\rangle_{mn} = \textrm{exp}\left[ -\lambda_{mn} \right] M_{mn}$ \cite{Mahan2000}, where the exponent $\lambda_{mn}(T)$ depends on temperature $T$ and on the difference between the local $e$-ph couplings at the $m$ and $n$ WF sites,
\begin{gather}
\label{thermal_average_M_tilde_local}
\lambda_{mn} (T) = \frac{1}{N_{\Omega}}\sum_{\textbf{Q}}
        \left( N_{\textbf{Q}}(T) + \frac{1}{2} \right)
        \big| g_{\textbf{Q}mm} - g_{\textbf{Q}nn} \big|^2,
\end{gather}
and on the phonon thermal occupation factor $N_{\textbf{Q}}(T)$. In this work, the quantity $\lambda_{mn}$ is computed using \textit{ab initio} $e$-ph coupling constants $g_{\textbf{Q}mm}$, paying attention to converge the Brillouin zone integral in Eq.~(\ref{thermal_average_M_tilde_local}). The diagonal part of $\lambda_{mn}$ is identically zero, which makes $\textrm{exp}[-\lambda_{mm}]=1$ for all sites $m$. The off-diagonal part of $\textrm{exp}[-\lambda_{mn}]$ is orders of magnitude smaller than unity (typically of order $10^{-2}$ to $10^{-10}$ at 300 K), as we verify explicitly in our numerical calculations. Thus we have
\begin{align}
\label{delta_identity}
\textrm{exp}[-\lambda_{mn}] \approx \delta_{mn}.
\end{align}
Substituting Eqs.~(\ref{ansatz}) to (\ref{delta_identity}) into Eq.~(\ref{polaron_parameters}), we derive the central equations for the polaron hopping strength $E_{mn}$ and pl-ph coupling $G_{\textbf{Q}mn}$:
\begin{gather}
\label{polaron_onsite_energy}
E_{mn}=\Big( \varepsilon_{mm}-\frac{1}{N_{\Omega}}
\sum_{\textbf{Q}}\omega_{\textbf{Q}}
\big| g_{\textbf{Q}mm} \big|^2 \Big) \delta_{mn}
,\\
G_{\textbf{Q}mn} = 0.\nonumber
\end{gather}
The first equation implies that the operators $a^{\dagger}_m$ in the polaron Hamiltonian, Eq.~(\ref{H_tilde_ta}), create a self-localized polaron because inter-site hopping is negligible due to the vanishing off-diagonal $E_{mn}$ elements. The second line implies that this small polaron state is decoupled from all phonon modes as $G_{\textbf{Q}mn} = 0$. The on-site polaron energy $E_{mm}$ is the sum of the electronic energy $\varepsilon_{mm}$ of the corresponding WF and the potential energy decrease due to the lattice distortion, analogous to the CHO case [compare the second terms in Eqs.~(\ref{transformed_H_CHO}) and (\ref{polaron_onsite_energy})].\\ 
%
% Figure 1
%
\begin{figure}[!b]
\includegraphics[width=0.99\linewidth]{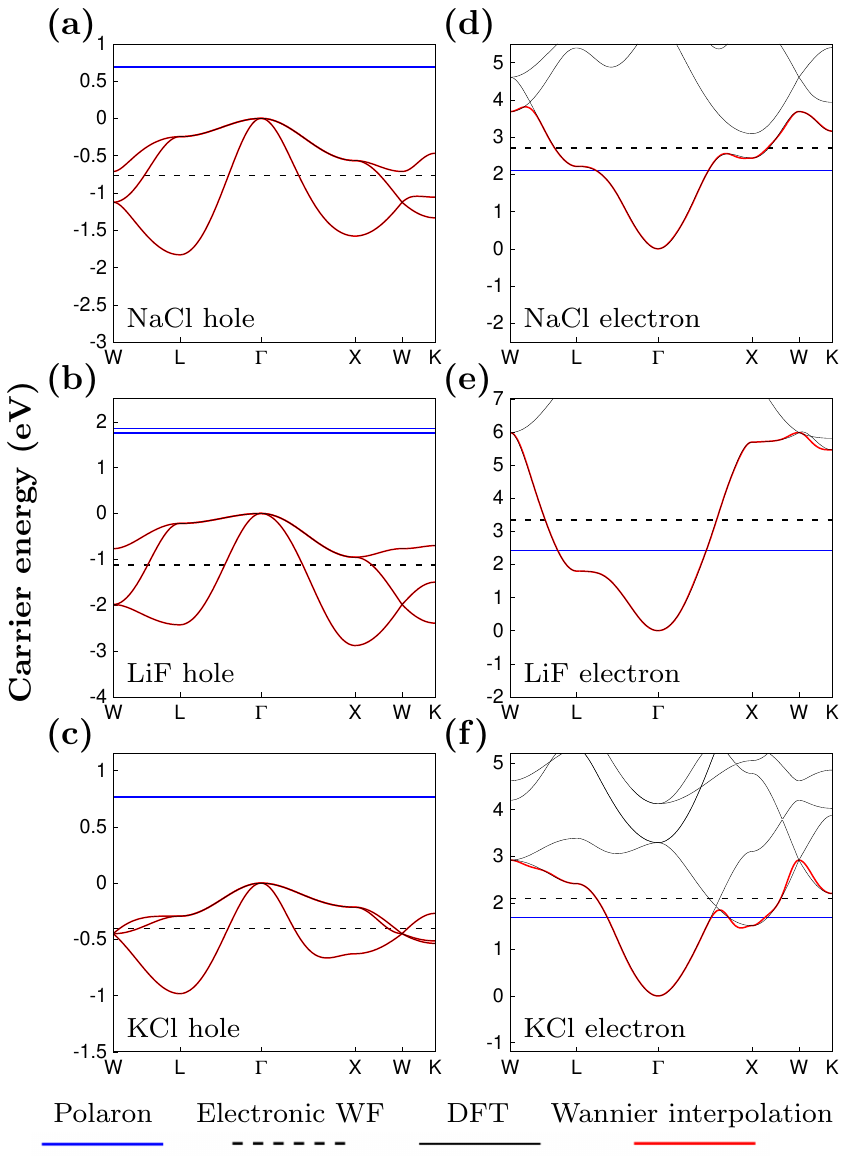}
\caption{Calculated polaron energy for holes in (a) NaCl, (b) LiF and (c) KCl, and electrons in (d) NaCl, (e) LiF and (f) KCl. Blue lines are the polaron on-site energies $E_{mm}$, and dashed black lines are WF energies $\varepsilon_{mm}$, in Eq.~(\ref{polaron_onsite_energy}). The solid black curves are the DFT band structure and the red curves are the Wannier interpolated bands, whose number equals the number of WFs employed in the calculation. The energy zero is set to either the CBM or VBM, respectively when electron or hole carriers are considered.}
\label{fig:polaron_energy_plot}
\end{figure}
\indent
Whether or not a small polaron forms depends on the competition of two terms, the potential energy decrease due to the lattice distortion and the kinetic energy increase from localizing a Bloch state. If the on-site polaron energy $E_{mm}$ is lower than the energy of the conduction band minimum (CBM) for an electron carrier, or higher than the valence band maximum (VBM) for a hole carrier, then the self-localized polaron is energetically more favorable than a delocalized Bloch state. 
In this scenario, the electron or hole quasiparticle forms a small polaron and becomes self-trapped by the lattice distortion; the polaron formation energy is thus the difference between the polaron energy $E_{mm}$ and the respective band edge. The physical insight provided by Eq.~(\ref{polaron_onsite_energy}) is that a material with less dispersive electronic bands, in which $\varepsilon_{mm}$ is closer to the band edge, and stronger on-site $e$-ph coupling $g_{\textbf{Q}mm}$ (and thus greater potential energy decrease) is more likely to host a small polaron.\\
\indent %created by the operator $a^{\dagger}_m$
The small polaron wavefunction has rarely been discussed in the canonical transformation treatment. While Eq.~(\ref{polaron_onsite_energy}) gives the polaron energy for an electron in a given WF, the choice of a WF is not unique $-$ different WFs will result in slightly different lattice distortions and polaron energies, the most stable state corresponding to the WF minimizing the polaron energy. In the following, we use a maximally localized WF as a trial wavefunction and compute its polaron energy. If the resulting small polaron is stable, as determined by comparing the electron or hole polaron energy with the respective band edge, then our approach provides a sufficient condition for concluding that a small polaron forms in the material, as well as an approximate polaron wavefunction.\\
%Systematic minimization of the polaron energy, which is left for future work, could more accurately determine the polaron wavefunction starting from an initial WF guess.\\
% though it cannot rule out the existence of a small polaron if that does not exceed the respective band edge.\\
%
\indent
We carry out DFT calculations using the {\sc Quantum ESPRESSO} code \cite{QE-2009} with a plane-wave basis set, employing norm-conserving pseudopotentials \cite{Troullier1991} from Pseudo Dojo \cite{VANSETTEN2018} and the Perdew-Burke-Ernzerhof generalized gradient approximation \cite{Perdew1996}. A kinetic energy cutoff of 100 Ry, an $8 \times 8 \times 8$ \textbf{k}-point grid and relaxed lattice parameters are used in all DFT calculations. %for all materials. 
We use density functional perturbation theory \cite{Giannozzi2001} to compute phonon frequencies and eigenvectors on a coarse $8 \times 8 \times 8$ \textbf{q}-point grid for all materials except Na$_2$O$_2$, for which we use a $4 \times 4 \times 4$ \textbf{q}-point grid. The $e$-ph coupling constants $g_{mn\nu}$(\textbf{k}, \textbf{q}) are obtained on coarse grids and transformed to Wannier basis coupling constants $g_{\textbf{Q}mn}$ using the {\sc Perturbo} code \cite{Perturbo2020}, with WFs generated from {\sc Wannier90}~\cite{Marzari2014}. 
%\\
%
% Figure 2
%
\begin{figure}[!b]
\includegraphics[width=0.99\linewidth]{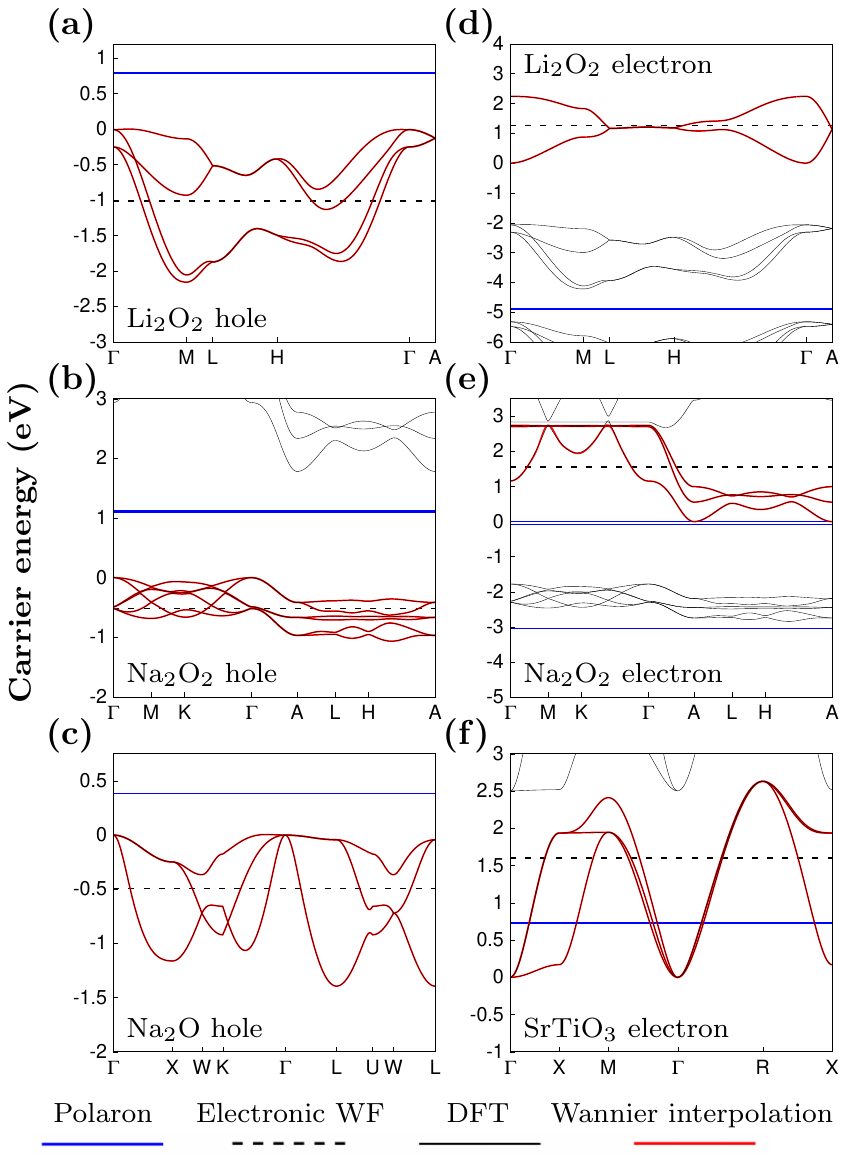}
\caption{Calculated polaron energy for holes in (a) Li$_2$O$_2$, (b) Na$_2$O$_2$ and (c) Na$_2$O, and electrons in (d) Li$_2$O$_2$, (d) Na$_2$O$_2$ and (f) SrTiO$_3$. The lines and their color code have the same meaning as in Fig. \ref{fig:polaron_energy_plot}.}
\label{fig:polaron_energy_plot_2}
\end{figure}
%
%
%
%\indent
Computing the polaron formation energy with Eq.~(\ref{polaron_onsite_energy}) only requires wannierizing one or more bands and computing the potential energy decrease term, which has a small computational cost equal to computing an $e$-ph scattering rate~\cite{Perturbo2020}. 
Before calculating the polaron energy, we first numerically verify that the identity in Eq.~(\ref{delta_identity}) is satisfied. 
%by carrying out the integration in Eq. (\ref{thermal_average_M_tilde_local}). 
We then obtain the on-site polaron energy, $E_{mm}$ in Eq.~(\ref{polaron_onsite_energy}), carrying out the Brillouin zone integral via Monte Carlo integration with 1 million random \textbf{q} points drawn from a Cauchy distribution. All materials investigated in this work have strongly polar bonds and dominant Fr\"{o}hlich $e$-ph coupling with the longitudinal optical modes~\cite{Zhou2016, Jhalani2020}. The temperature is set to 300 K in all calculations.\\
\indent
Figure \ref{fig:polaron_energy_plot} shows the computed polaron energy in three alkali halides, NaCl, LiF and KCl, for both electron and hole polaron states. Our formalism predicts that holes in these three materials form a self-localized small polaron, in agreement with experiments~\cite{Castner1957}, because the computed polaron energies $E_{mm}$ are above the VBM, as shown in Fig.~\ref{fig:polaron_energy_plot}(a)$-$(c). Electrons in these materials, on the other hand, are not expected to self trap $-$ as the conduction band in alkali halides is $s$-like and therefore more dispersive than the $p$-like valence band, the potential energy decrease due to the lattice distortion cannot outweigh the increase in kinetic energy for localizing the electronic state. Consistent with this intuition, our results for electrons in NaCl, LiF and KCl, shown in Fig.~\ref{fig:polaron_energy_plot}(d)$-$(f), conclude that electrons in these materials do not form a self-trapped polaron, as is seen by the fact that the polaron energy is above the CBM. Experiments in alkali halides similarly found no evidence of electron polarons down to 5 K temperature~\cite{Duerig1952}.\\
% searching self-trapped electron states
%
% Figure 3
%
\begin{figure}[!b]
\includegraphics[scale=0.87]{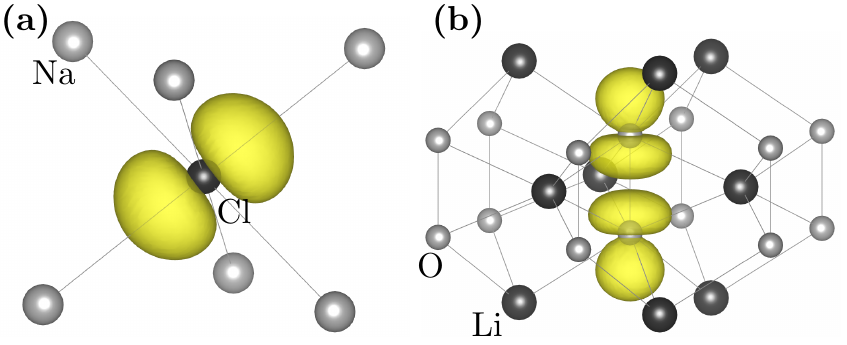}
\caption{The square of the trial polaron wavefunction for the (a) hole polaron in NaCl and (b) electron polaron in Li$_2$O$_2$.}
\label{fig:wannier_function}
\end{figure}
%
%
%Figures \ref{fig:polaron_energy_plot_2}(a)$-$(e) strong $e$-ph coupling and 
\hspace{3pt}Figure~\ref{fig:polaron_energy_plot_2} shows the calculated electron and hole polaron energies in three alkali metal oxides and peroxides, Na$_2$O$_2$, Li$_2$O$_2$ and Na$_2$O. The nature of the charge carriers in these materials is important for application to novel battery technologies, where the low electrical conductivity hampers device performance and is commonly attributed to the presence of small polarons~\cite{Kang2012}. Our results in Fig.~\ref{fig:polaron_energy_plot_2} unambiguously demonstrate that both electrons and holes in these materials form self-localized small polarons with formation energies greater than 0.5$-$1 eV, warranting further investigation of their electrical transport properties. 
The hole polaron wavefunction in NaCl and the electron polaron wavefunction in Li$_2$O$_2$ are shown in Fig.~\ref{fig:wannier_function}(a) and (b), respectively, highlighting their localized nature.\\
\indent
The last case study we examine is cubic SrTiO$_3$ perovskite, whose electron mobility near room temperature exhibits a power law that can be attributed to a transport regime governed by large 
(non-self localized) polarons~\cite{Zhou2019}. We investigate small polaron formation in cubic SrTiO$_3$, using accurate electronic bandstructure, phonon dispersions and $e$-ph interactions from our previous work~\cite{Zhou2018, Zhou2019} as a starting point for the polaron calculation. As shown in Fig.~\ref{fig:polaron_energy_plot_2}(f), we find a polaron energy significantly higher than the CBM, clearly showing that for electrons in SrTiO$_3$ it is energetically unfavorable to self-localize and form a small polaron state. Note that this finding does not conflict with the existence of localized electronic states due to oxygen vacancies \cite{Janotti2014, Crespillo2018} as our approach focuses on self-localized electronic states in the pristine crystal.\\
% in-gap inherently 
%
%
\indent
The formalism presented in this work leaves room for various extensions. One is minimizing the polaron energy over the space of possible trial WFs, leading to a refinement of the polaron formation energy and wavefunction. Mode-resolved analysis of the potential energy decrease is also possible, and allows one to infer which phonon modes contribute to small polaron formation. 
In addition, treating $e$-ph interactions in materials with open-shell $d$ or $f$ electrons, for example using the DFT+U approach, is an important future extension for studies of small polaron effects in transition metal oxides~\cite{Setvin2014, Freytag2016}. Our approach also forms the basis for charge transport calculations in the polaron hopping regime, for example using the Kubo formula~\cite{Lang1962, Munn1980, Tim-PRX}, and for studies of the transition from bandlike to polaron hopping transport~\cite{Schein1978, Bogomolov1967, Bottger1976}. Both topics are pristine territory for first-principles calculations. 
%\\
%
%
%
%\indent %and practical 
Finally, there are important conceptual differences between our approach and a recently proposed momentum-space formalism to treat small polarons~\citep{Sio201901, Sio201902}, as we will discuss elsewhere.\\
\indent
In summary, we developed a computationally efficient approach to predict the formation of self-localized small polarons, and made it available in our open source \mbox{{\sc Perturbo}} code. 
Our formalism combines \textit{ab initio} $e$-ph interactions with an extension of small polaron theory. %deriving a predictive framework to investigate the formation of self-localized polaron states. 
Its computational cost is a minimal overhead to a DFT calculation on a unit cell, allowing one to rapidly scan many materials. %It is ideally suitable for high-throughput investigations of polaron effects in materials. %, including alkali halides and metal oxides of technological relevance.
Besides providing a convenient atomistic approach for small polaron studies, our method is a starting point for developing transport calculations in the polaron hopping regime.

This work was supported by the Air Force Office of Scientific Research through the Young Investigator Program, Grant FA9550-18-1-0280. J.-J. Z. was supported by the Joint Center for Artificial Photosynthesis, a DOE Energy Innovation Hub, supported through the Office of Science of the U.S. Department of Energy under Award No. DE-SC0004993. H.-Y. C. acknowledges support by the J. Yang Fellowship. This research used resources of the National Energy Research Scientific Computing Center, a DOE Office of Science User Facility supported by the Office of Science of the US Department of Energy under Contract No. DE-AC02-05CH11231.

\bibliographystyle{apsrev4-2}
\bibliography{polaron}
\end{document}